# Direct observation of pore collapse and tensile stress generation on pore wall due to salt crystallization

A.Naillon[1,2,3], P.Joseph[2], M.Prat[1*]

[1]*Institut de Mécanique des Fluides de Toulouse (IMFT),Université de Toulouse, CNRS, Toulouse, France*
[2]*LAAS-CNRS, Université de Toulouse, CNRS, Toulouse, France*
[3]*Univ. Grenoble Alpes, CNRS, Grenoble INP, LRP, 38000 Grenoble, France*

**Abstract** – The generation of stress in a pore due to salt crystallization is generally analysed as a compressive stress generation mechanism using the concept of crystallization pressure. We report on a completely different stress generation mechanism. In contrast with the classical picture where the crystal pushes the pore wall, the crystal growth leads to the generation of a local tensile stress. This tensile stress occurs next to a region where a compressive stress is generated, thus inducing also shear stresses. The tensile stress generation is attributed to capillary effects in the thin film confined between the crystal and the pore wall. These findings are obtained from direct optical observations in model pores where the tensile stress generation results in the collapse of the pore region located between the crystal and the pore dead-end. The experiments also reveal other interesting phenomena, such as the hyperslow drying in PDMS channels or the asymmetrical growth of the crystal during the collapse.

**Significant statements** – Porous materials, including stone, brick or concrete can be damaged when salt crystals precipitate in theirs pores. The stress generation on the pore wall due to the growth of a single crystal is studied from direct optical observations in a model dead–end pore. The experiment shows that the crystal growth can lead not only to the generation of a compressive stress as expected from previous works but also to the pore collapse and to the generation of a tensile stress. The latter are attributed to capillary effects in the thin film confined between the crystal and the pore wall. The study also analyses various phenomena characterizing the confined crystal growth and the drying process in the model pore.

---

* Corresponding author : <mprat@imft.fr>, +33 (0)5 34 32 28 83

During evaporation from a porous media containing dissolved salts, the salt concentration increases and can reach a sufficient concentration for salt crystals to form. As reported for example in [1-3], the presence of the ions can dramatically change the drying kinetics owing to the formation of salt crust or pore clogging. Still more importantly in relation with civil engineering and cultural heritage conservation issues, e.g. [4], crystal formation can cause severe damages and cracks in porous materials, e.g. [5-7]; sometimes leading to a complete destruction [8]. The stress generation mechanism leading to damages is generally associated with the concept of crystallization pressure, e.g. [9-11] and references therein. The latter can be expressed for sufficiently large crystals of sodium chloride (>1μm) as (only NaCl is considered throughout this paper),

$$P_c = \frac{2RT}{V_m}\left(\ln S + \ln \frac{\gamma_\pm}{\gamma_{\pm,0}}\right), \qquad (1)$$

where $R$ is the ideal gas constant, $T$ is the temperature, $V_m$ is the molar volume of the solid phase forming the crystal ($V_m$ = 27.02 cm$^3$/mol for NaCl), $\gamma_\pm$ is the ion mean activity coefficient. Index $_0$ refers to the reference state where the crystal is in equilibrium with the solution in the absence of stress applied on the crystal. The ratio $S = m/m_0$ is the supersaturation, where $m$ denotes the molality of the solution ($S$ = 1 when the crystal and the solution are in equilibrium in the reference state). According to Eq.(1), stress can be generated when the solution in contact with the crystal is supersaturated ($S$ > 1). Although supersaturation as high as 1.7 have been measured [12-14], the stress actually generated cannot be readily deduced from Eq.(1). As discussed in [11], what matters is not the supersaturation at the crystallization onset but the supersaturation when the crystal is about to clog the pore. The latter is generally much smaller owing to the local consumption of ions near the growing crystal during its growth. The net result however, when the conditions are met for the crystal to touch the wall with a sufficiently high supersaturation, is the generation of a compressive stress on the pore wall, meaning here that the crystal tends to "push" the wall, e.g. [11]. One can also refer to [10] for more details on how the crystal can push the wall through a thin liquid film confined between the crystal and the pore wall.

In what follows we describe a completely different mechanism of stress generation induced by the crystallization, leading, at least for our system, to stresses comparable in magnitude with the stresses due to crystallization pressure. In contrast with the classical case, however, the generated stress is not compressive but tensile and this results in the collapse of the dead-end section of the pore.

The study is based on observations in a microfluidic device where the crystallization is generated by evaporation of a NaCl aqueous solution confined in dead-end channels. In addition to tensile stress generation, the experiments reveal interesting phenomena such as a hyperslow evaporation kinetics, the crystallization induced acceleration of the receding meniscus and the preferential growth of crystal on the side of the dead–end section of the channel, thus shedding light on the rich physics of crystal growth at the pore scale.

**Experimental set-up**

The microfluidics device and its fabrication procedure have been presented in previous papers, e.g. [11] and [14] and therefore is only briefly described here (see however SI Appendix A for a Figure and additional details). Evaporation experiments of a saline aqueous solution are performed in dead-end square channels of 4.5×4.5μm² cross section surface area referred to as pore channels. The channel length is 200μm. The chips containing the channels are of PDMS and glass. The glass is used for the cover plate closing the PDMS channels. Salt solution is prepared with NaCl provided by Sigma

Aldrich© dissolved in deionized water. Unless otherwise mentioned, the initial molality of the solution is 1.89, the saturation in the reference state being 6.15 mol/kg (corresponding to mass fractions of 10% and 26.4 % respectively). Salt purity is ensured to be higher than 99.5%.

Experiments are performed on an inversed microscopy Zeiss Axio observer D1 working in transmission. The dead-end channels are filled with a salt solution of known concentration. Then, they are dried maintaining a nitrogen flow at their entrances during all the experiment.

Crystallization starts once a critical salt concentration is reached in the pore channel. An Andor Zyla SCMos camera is used to record the kinetics of evaporation with a frame rate of 2 seconds per image.

**Observations**

A meniscus forms and progressively recedes into each pore channel. During the evaporation of a salt solution, only water evaporates whereas the dissolved species remain trapped in solution. As a result, salt concentration increases up to reach a higher value than the equilibrium one. This metastable state lasts until the onset of crystallization. Once nucleation occurs, crystal grows consuming the ions in excess above the equilibrium concentration. At the same time, the solution continues to evaporate providing more salt for crystal growth. This sequence is illustrated in Fig.1.

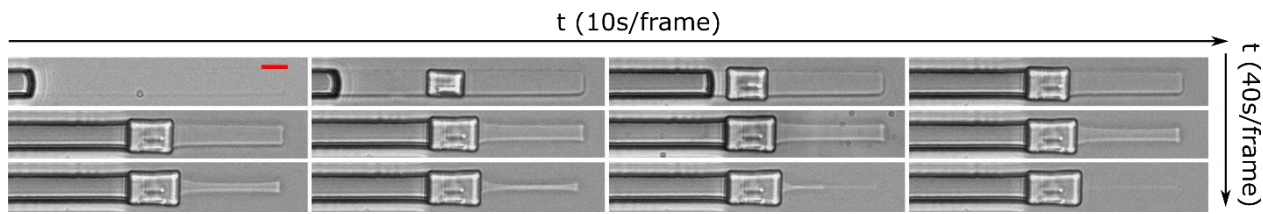

**Fig. 1.** *Channel deformation resulting from the growth of a single cluster. A crystal appears in the channel of 4.5× 4.5µm² cross-section surface area at some distance of the receding meniscus (visible on the left on the first three top images) and clogs it. The crystal growth first induces a positive deformation when the crystal touches the wall and continues to grow. The meniscus continues receding. When it reaches the crystal, the channel wall on the right of crystal starts to collapse. At the end, the channel on the right of the crystal is totally closed. The red scale bar represents 5µm.*

As can be seen, the formation of the crystal first leads to a positive deformation. This is better illustrated in Fig.2a, which shows that the crystal width is greater than the initial channel width $W_i$. For more details on the compressive stress generation, one can refer to [11] where a stress diagram summarizing the conditions leading to the stress generation is presented. Fig.1 also shows an unexpected phenomenon: the collapse of the channel on the dead–end side, i.e. the section of the channel located between the most advanced face of the crystal and the channel dead–end. As can be seen, the collapse is progressive and the crystal continues to grow during the collapse period. However, as better shown in Fig.2, the crystal growth takes place only on the right, i.e. where the crystal is in contact with the collapsing region. Note that the back face is defined as the crystal face on the side of the pore channel open end whereas the front face is the face of the crystal on the side of the pore channel dead-end, thus the most advanced face into the channel. The width of the collapsing region is the minimum width in the images of the collapsing region. It roughly corresponds to the width in the middle of the collapsing region. In contrast with the classical picture describing the effect of crystallization as pushing the pore walls, we observe here the generation of tensile stress pulling the walls toward the interior of the channel.

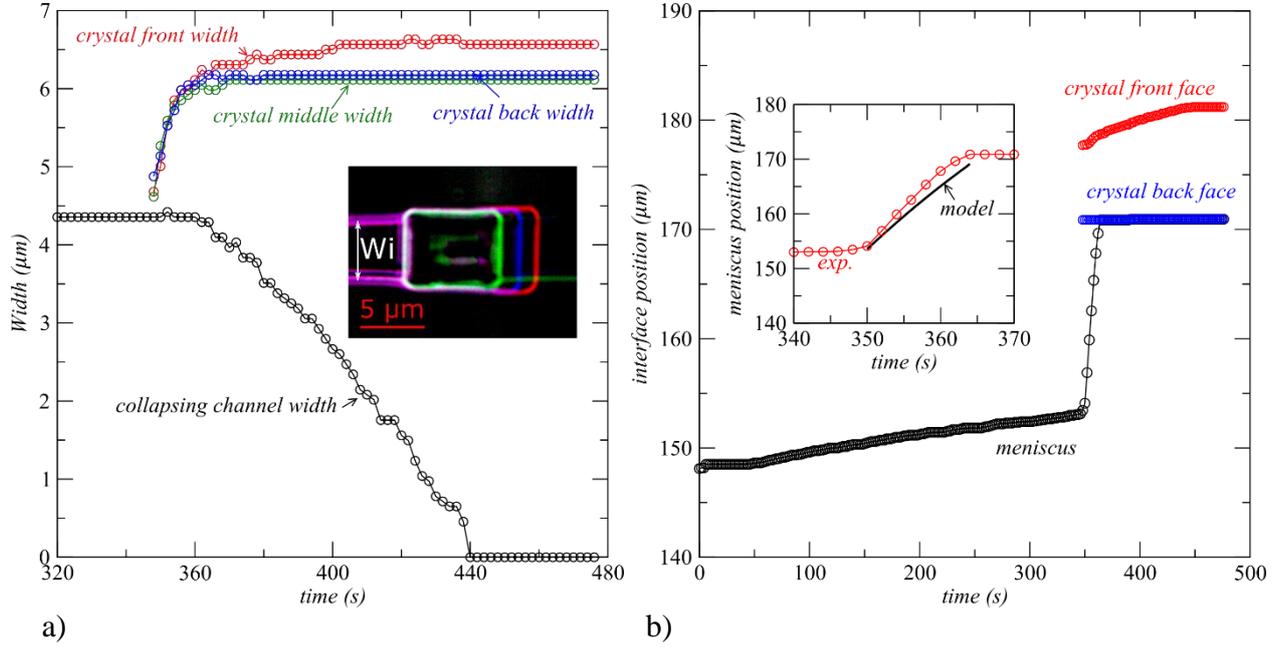

**Fig.2.** a) *Variation of channel width in collapsing region and crystal back and front face widths as a function of time. The inset shows superimposed images of the crystal contour at three different times. Growth occurs on the right side of the crystal, i.e. on the side of channel collapsing region; $W_i$ is the initial channel width.* b) *Variation of meniscus, crystal back face and crystal front face positions as a function of time. The position is the distance from the pore channel open end. The inset shows the comparison between model (see SI Appendix C) and experimental results in the acceleration period of the meniscus.*

**Hyperslow drying**
In addition to this unexpected collapse phenomenon, the experiments reveal several other interesting phenomena. First, as illustrated in Fig.3, the receding meniscus kinetics is much slower than expected if one assumes that the evaporation should follow the classical Stefan's tube evaporation kinetics [15]. The latter predicts that the receding meniscus position in the tube (distance between the channel open end and the meniscus) varies as $\sqrt{\frac{2D_v \rho_{vs}}{\rho_\ell}t}$ where $t$ is the time, $D_v$ is the molecular diffusion of the water vapor in nitrogen, $\rho_\ell$ is the solution density, $\rho_{vs}$ is the water vapor concentration at the meniscus surface. To obtain the result shown in Fig.3, we have taken for simplicity the value of $\rho_{vs}$ corresponding to a sodium chloride saturated solution. Note that the experimental results shown in Fig. 3 are for an initial salt fraction of 20% (and not 10% as for the other results shown in the papers). This is just because the data on the meniscus position are available right from the beginning for this particular experiment. The experiments have been performed several times with different configurations (geometry and initial concentration) and the hyperslow drying has always been observed.

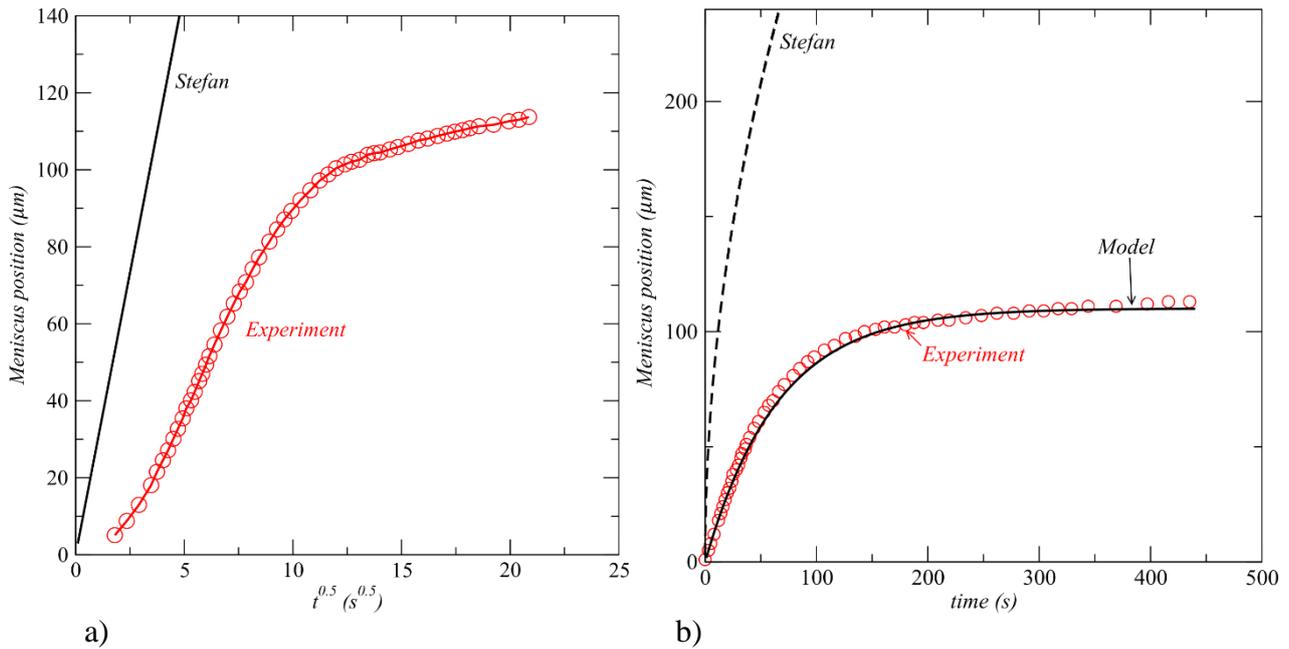

**Fig.3.** *Hyperslow evaporation kinetics in the channel. a) Meniscus position as a function of the square root of time. The curve labelled "Stefan" corresponds to the classical diffusion controlled evaporation kinetics in a straight tube. b) Comparison between the experiment and the pervaporation – condensation model described in SI Appendix B.*

To explain the hyperslow evaporation depicted in Fig.3, it should be recalled that water can actually migrate into PDMS, [16-20]. Since the chip is first invaded by the solution during several minutes before the evaporation starts, the PDMS is actually saturated with water, at least near the pore channel and supply channel walls. The simple model taking into the pervaporation process presented in SI Appendix B leads to the comparison with the experimental data shown in Fig.3. As illustrated in Fig.3, it leads to consistent results with the experiment. However, one might wonder why the meniscus motion is (much) slower than predicted by Stefan's model. One might think that the pervaporation process acts in addition to the vapor diffusion transport within the channel. Accordingly, the meniscus motion should be faster than predicted by Stefan's model. The explanation is the following. Due to the presence of water in the channel PDMS walls, water is actually transferred from the wall into the gas phase in the channel (see a schematic of the process in SI Appendix B). Thus, the vapor concentration in the gas in the channel is expected to be close to the saturation vapor concentration, $\rho_v \sim \rho_{vs}$. As a result, the vapor concentration gradient along the channel is much less than in the classical Stefan's situation. In other words, the vapor diffusive transport in the channel is expected to be negligible. In summary: i) the pervaporation process and the humidity inside the PDMS are responsible for the very slow meniscus motion in the channel, ii) the meniscus motion is very slow at the onset of crystallization (which occurs right at the end of the period shown in Fig.3b).

**Meniscus sudden acceleration**

As shown in Fig.2b, the meniscus suddenly and strongly accelerates when $t \approx 350$ s. As can be seen from Fig.2b, this strong acceleration (the change in the slope of the meniscus position curve in Fig.2b is by a factor of about 70) is concomitant with the crystal growth in the channel. This acceleration is explained by two phenomena. The less important one is related to the deformation of the channel induced by the crystal growth near the crystal back face. A simple volume conservation argument implies the acceleration of the meniscus because of the channel cross section surface area increase due to the channel deformation by the crystal. According to the model presented in SI Appendix C, the second phenomenon is more important. Owing to the greater density of the crystal compared to the salt concentration in the solution, the growth of the crystal induces a liquid flow in direction of the growing crystal interface [11]. This flow induces in turn the acceleration of the meniscus. As depicted in the inset in Fig.2b, a simple model taking into account both phenomena (see SI Appendix C) leads to a good agreement with the experimental data.

**Crystal front face longitudinal growth**

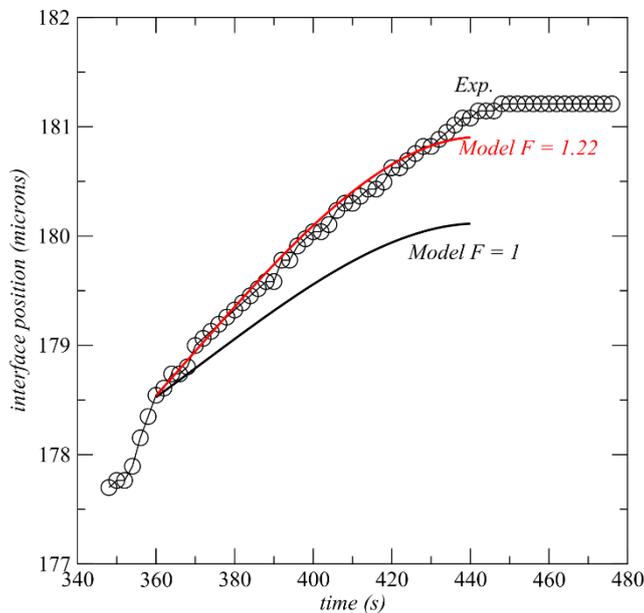

**Fig.4.** *Crystal front face position as a function of time. Comparison between the experimental data and a model based on the assumption that the crystal front face growth is due to the precipitation of ions contained in the collapsing liquid plug. Parameter F is a shape factor characterizing the shape of the collapsing region cross section surface area (see SI Appendix D)*

To explain now the growth of the crystal at the front and not at the back depicted in Fig.2, we first note that the back face transverse growth stops when the meniscus reaches the crystal (compare Figs. 2a and 2b). This indicates that this growth should correspond to the precipitation of the ions contained in the liquid plug on the left of the crystal, i.e. the liquid plug between the crystal back face and the receding meniscus. The fact that this growth stops when this liquid plug disappears is an indication that little ions, if any, are transported from the channel collapsing region up to the crystal back face. This is also consistent with the fact that the evaporation at the meniscus is too weak for inducing a noticeable liquid flow in the direction of the channel entrance during the time of crystal growth. The noticeable growth of the front face is analyzed similarly. This growth must correspond to the precipitation of the ions contained in the channel collapsing region. Assuming that the salt concentration in the collapsing region is the equilibrium concentration (which is consistent with the results reported in [13-14] showing that the ions in excess at the crystallization onset are very rapidly consumed), a simple mass conservation model based on this assumption (see SI Appendix D) leads to the comparison depicted in Fig.4. The favorable comparison between this model and the experiment supports the proposed analysis.

**Crystal front face transverse growth**

Then, we have to explain why only the advancing region of the crystal grows transversally and pushes the wall and not the region of the crystal located further away from the advancing crystal face. We first note that the disappearance of the liquid in the collapsing region cannot be explained by the evaporation in the region of the crystal back face since the evaporation is, as discussed previously, quite low at the receding meniscus just before the crystal grows and the collapse occurs. Noting that the halite (the crystallized form of sodium chloride) is anhydrous, the conclusion is that water leaves the collapsing region through pervaporation of water through the PDMS, e.g. [16-20]. Thus, as schematically illustrated in Fig.5, the picture is that water leaves the collapsing region by pervaporation through the PDMS while ions precipitate on the crystal. Thus, the collapse kinetics is controlled by the pervaporation process. The pervaporation velocity $v_{pe}$ is estimated in SI appendix E as being $v_{pe} = 2.3 \times 10^{-8}$ m/s.

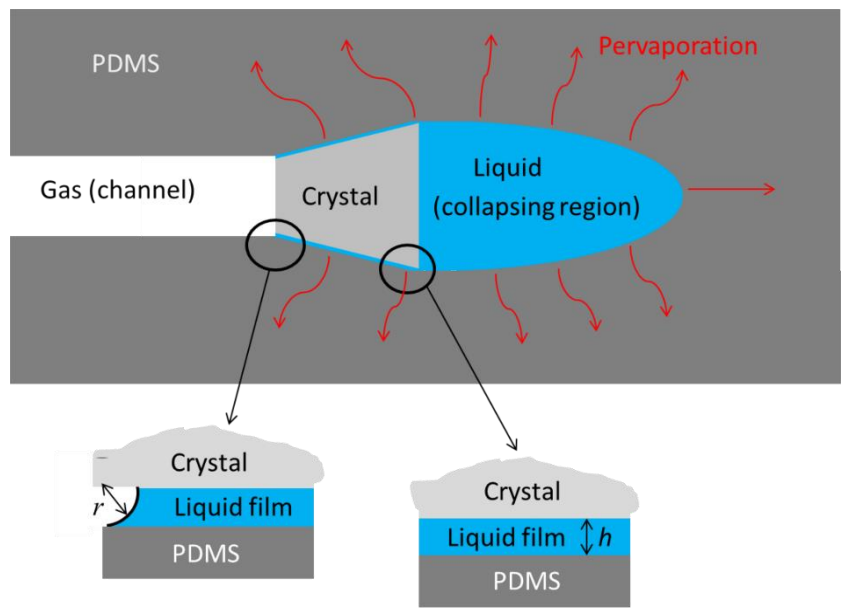

**Fig.5.** *Schematic of channel collapse situation. The red arrows represent the pervaporation process.*

Also, we consider the somewhat classical picture, e.g. [5,10,21,22], where a thin liquid film of thickness *h* is present between the crystal and the channel wall. This film is necessary for the transverse growth of the crystal since ions must access to the crystal surface for making it grow. The film is sketched in Fig. 5. Then the model of the ion transport in the film presented in SI Appendix F leads to the height-averaged ion mass fraction profiles depicted in Fig.6. According to the classical diffusion reaction theory (DRT) [23], the crystal growth is analyzed as a reaction process during which ions fit in the crystal lattice. The latter is expressed as

$$w_{cr} = \frac{k_r \rho_\ell}{\rho_{cr}} (C_i - C_{eq}) \quad (2)$$

where $w_{cr}$ (m/s) is the local crystal growth rate; $k_r$ (m/s) is the reaction (precipitation) coefficient, $C_i$ (kg/m$^3$) is the ion mass fraction at the crystal surface, $C_{eq}$ is the ion mass fraction at equilibrium and $\rho_{cr}$ is the crystal density (kg/m$^3$). Thus the ion mass fraction in the solution must be (slightly) greater than $C_{eq}$ for the crystal to grow. The profiles depicted in Fig.6 are thus fully consistent with the experiments since they indicate that the growth occurs only in the region of the film located in the very close vicinity of the front face of the crystal. Further away from the film entrance $C \sim C_{eq}$, which is consistent with the observation of no transverse growth away from the crystal front face (Fig.2a). Since the experimental observation indicates that the transverse growth is quite localized at the edge of the crystal front face the results plotted in Fig.6 suggests that the film thickness is closer to 10 nm than 100 nm. This is consistent with the thicknesses reported in [22] where another interesting situation where transport phenomena in the film control the growth of a confined crystal is analyzed. As explained in SI Appendix F the ion mass fraction $C_{L_c}$ at the entrance of the film ($x = L_c(t)$) is estimated from the measured growth rate of the crystal and Eq.(2). This yields $C_{L_c}/C_{eq}$ =1.00013.

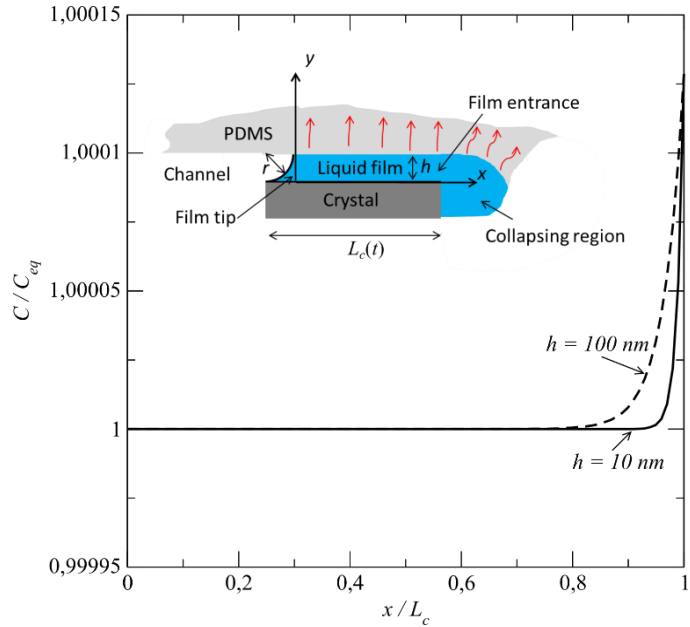

**Fig.6.** *Ion mass fraction distribution along the film for two thicknesses of the film. The inset shows a schematic of the thin film between crystal and wall with notations. For the sake of clarity, the ratio h / $L_c$ in the inset is not at scale (h ~ 10 – 100 nm, $L_c$ ~10 μm).*

**Mechanical considerations**

Another reasonably consistent aspect lies in the value of the crystallization pressure corresponding to the estimate of the ions in excess at the film entrance depicted in Fig.6 ($C_{L_c}/C_{eq}$ =1.00013). To this end, we consider the mechanical equilibrium in the film region. Assuming negligible wall and crystal curvature effects, this (quasi-static) mechanical equilibrium can be expressed as [21]

$$\sigma = -p_s - p_d \qquad (3)$$

where σ is the normal stress on the pore wall (negative when compressing the wall using the same convention as in [21]), $p_s$ is the pressure in the solution and $p_d$ is the disjoining pressure [24]. According to the analysis presented in [21], $p_d$~$P_c$, where $P_c$ is the crystallization pressure (Eq.(1)). Note that only equilibrium situations are considered in the analysis presented in [21]. We thus assume that this analysis is still acceptable under non-equilibrium conditions, i.e. during the crystal growth. The pressure $p_s$ in the solution is expressed as $p_{atm}$ - $P_{cap}$, where $p_{atm}$ is the atmospheric pressure (pressure in the gas phase) and $P_{cap}$ is the capillary pressure (the pressure jump between the liquid and gas phase through the meniscus sketched in Fig.5). Applying Laplace's law with a contact angle ~90° on the PDMS wall and a zero contact angle on the crystal [25] leads to express $p_s$ as $p_s = p_{atm} - \frac{\gamma}{r}$ where $r$ is the curvature radius depicted in Fig.5 (γ is the surface tension, γ ≈ 83 × $10^{-3}$ N/m for a saturated NaCl aqueous solution). The meniscus curvature in the film plane is neglected since the film thickness is much smaller that the channel width. Thus, Eq.(3) is finally expressed as,

$$\sigma + p_{atm} = P_{cap} - P_c = \frac{\gamma}{r} - P_c \qquad (4)$$

The computations reported in [11] indicate that the compressive stress necessary to observe a channel deformation of 5 μm is about -0.5 MPa. In the present experiment (Fig.2a), the channel deformation is less ~ 1.3 μm based on the variation of the crystal back width in Fig.2a. As a result, the compressive stress to observe this deformation is expected to be on the order of -0.5 x 1.3/5 ~ 0.13 MPa. As computed in SI Appendix G, it is expected that the capillary pressure is on the order of 0.5 MPa. From Eq.(4), it is thus expected that $P_c$ ~ 0.53 MPa when no further transverse growth of the crystal occurs during the collapse. From Eq.(1), this corresponds to $C_{eq}$ / $C_{sat}$ ~ 1.0052 ($C_{eq}$ is the ion mass fraction in the film away from the crystal front face, where no additional crystal growth occurs, see [13] for the relation between ion mass fraction and molality; $C_{sat}$ is the solubility, i.e. the equilibrium ion mass fraction in the solution in the absence of stress applied on the crystal, i.e. in the reference state). In the region of the film adjacent to the crystal front face where the transverse deformation occurs the compressive stress, and thus the crystallization pressure, must be greater so as to cause the additional transverse deformation. Assuming an elastic deformation, the compressive stress causing the additional transverse deformation is proportional to the additional displacement, hence $\sigma(L_c(t)) = \sigma_{eq} e(L_c(t))/e_{eq}$ where, as depicted in Fig.6, $L_c$ is the length of the crystal. The displacement $e$ is defined as $e = W - W_i$ where $W_i$ is the initial width of the channel and $W$ is the width of the channel after deformation. With $e(L_c(t))$ ~ 1.69 μm and $e_{eq}$ ~ 1.3 μm (from the data shown in Fig.2a), this gives $\sigma(L_c(t)) \sim -0.17$ MPa. According to Eqs.(4) and (1), this corresponds to $P_c \sim 0.57$ MPa. According to Eq.(1), such a crystallization pressure corresponds to an ion mass fraction ratio of $C(L_c(t)) / C_{sat} \sim 1.0055$, leading finally to $C(L_c(t)) / C_{eq} \sim 1.0003$. This is greater but close to the ion mass fraction computed from the film model (Fig.6) indicating that $C(L_c(t)) / C_{eq} \sim 1.00013$ at the entrance of the film where ions in excess are necessary for generating the extra stress and the transverse growth of the crystal font face. Based on the approximations made to obtain the various estimates, we conclude that the estimate of the supersaturation at the entrance of the film obtained from the mechanical considerations is consistent with the supersaturation obtained from the film model. Nevertheless, more refined analyses, probably implying detailed numerical simulations (for instance in the spirit of the work presented in [26]), are desirable to reach still more firm conclusions.

**Collapse mechanism**
Then we are left with the explanation for the collapse itself. Since the liquid is not replaced by gas in the collapsing region, the crystal region acts as a barrier preventing the gas to reach the liquid region located between the crystal and the channel dead-end. Also, we note that no bubble formation is observed in the collapsing region. Based on the elastic modulus $E$ of PDMS ($E = 1.2$ MPa, see SI of [11]), and assuming purely elastic deformation, the numerical computation on the collapse presented in SI Appendix G, indicates that a negative pressure on the order of -5 bars (-0.5 MPA) is sufficient to cause the observed collapse. Consistently with the observation, this is much less than the negative pressure required for the formation of a bubble by cavitation (~ - 9 MPa according to [27]).
As sketched in Fig.5 and Fig.6 and discussed above, a meniscus must be present at the film tip on the side of the crystal back face when the liquid plug on the left of the crystal in Fig.1 disappears. It is surmised that the curvature of this meniscus adjusts in response of the pressure decrease of the solution induced in the collapsing region by the pervaporation. For $p_s - P_{atm} \sim$ -5 bars, i.e. on the order of magnitude of the negative pressure in the solution to observe the collapse, applying Laplace's law, i.e. $p_S = p_{atm} - \frac{\gamma}{r}$, gives $r \approx$ 170 nm. This curvature radius is greater than the expected thin film thickness $h$ [22]. For this reason, the liquid – gas interface must remain stuck at the entrance of the liquid film on the back of the crystal during the collapse.
Then, a simple idea is to consider that the liquid mass loss by pervaporation in the collapsing region induces the increase in the curvature of the liquid – gas interface. As a result, the pressure in the solution decreases, i.e. is more and more negative. Thus, the collapse would result from the combination of pervaporation and capillary effects. However, it has been shown, e.g. [18], that significant negative pressures can also be induced by the pervaporation process in a liquid pocket surrounded by PDMS. In our experiments, this would correspond to the situation where the crystal is in direct contact with the PDMS walls so as to hydraulically isolate the collapsing liquid plug. In other terms, this situation is only possible if one considers that the liquid film between the crystal and the PDMS disappears. For instance, one might consider that the air-liquid interface moves between the PDMS and the crystal. However, this would mean curvature radii of the order of the film thickness and thus capillary pressures not consistent, i.e. much too big, with the pressure levels corresponding to the collapse. Also the presence of the liquid film is necessary to explain the positive deformation of the channel in the crystal region before the collapse occurs and to explain that the crystal continues to grow transversally during the collapse, on the side of the collapsing plug. This is a clear indication that a liquid film must be present between the crystal and the walls in this region during the collapse.

**After the collapse**

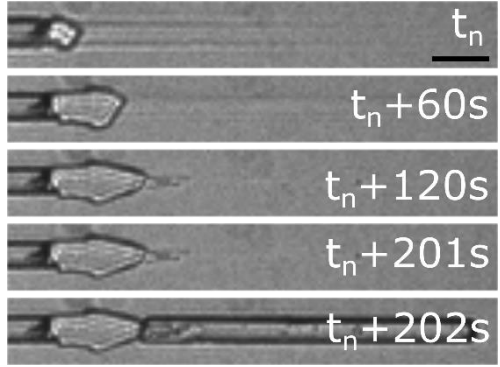

**Fig. 7.** *Example of channel collapse and instantaneous re-opening at the end of pervaporation. The re-opening occurs a few seconds after the end of collapse. Black scale bar is 10 µm.*

Also an additional interesting phenomenon can be observed after the collapse but not always. As illustrated in Fig.7, it can happen that the channel reopens after the collapse. In the example shown in Fig.7, the reopening occurs about 80s after the end of collapse. This is consistent with the disappearance of the thin film due to pervaporation. The mass of solution in the film is $m_{film} = \rho_\ell L_c hW$ (considering only one lateral face of the crystal). The pervaporation rate is $J_{pe} = \rho_\ell v_{pe} W L_c$. Thus, a characteristic time for the disappearance of the film by pervaporation is $t = m_{film} / J_{pe} = h / v_{pe}$. With $v_{pe} \approx O(10^{-8}$ m/s) (see SI Appendix E), this gives $t \approx 1 - 10$ s  for $h \sim 10 - 100$ nm. This estimate of the film disappearance time is compatible with the observation. The greater value observed in the experiment might be due to the presence of small cavities at the surface of the crystals containing some extra liquid.  Thus, in this case, the reinvasion of the film region by the gas phase together with the end of the capillary effect due to the liquid disappearance would lead to the channel reopening. It also happens that this channel reopening phenomenon does not occur. In this case, the channel remains collapsed and is still so after several weeks. It is surmised that the salt can sometimes fully clogged the film regions forming a barrier between the gas phase in the channel on the left of the crystal in Fig.7 and the collapsed region (contrary to water, the ions cannot leave the film region). In other words, during the pervaporation of the film, the salt precipitation can sometimes clog the film region and sometimes forms a thin zone through which the gas can percolate. Some variabilities in the adhesion forces between the PDMS walls in contact or between PDMS and glass might also play a role in these observations.

**Discussions**
In summary, our experiments in model pores first confirm from direct optical observations that the growth of a crystal in a pore can generate a compressive stress on the pore wall, see [11] for more details. Much more unexpectedly, we have shown that tensile stresses can be also generated. In the case of our experiments, this led to the collapse of the region located between the crystal and the model pore dead-end. The tensile stress was attributed to the negative pressure in the solution induced by capillary effect generated at the tip of the thin liquid film confined between the crystal and the pore walls. It can be noted that the process actually leads to a shear stress generation since a compressive stress and a tensile stress are generated together in about the same region of the pore wall. Although the mechanism of water loss inducing the capillary effect was due to the pervaporation of water through the PDMS wall of the model pores, it is surmised that similar capillary effects can be generated by evaporation at the film tip in the more classical situation where the solid matrix of the porous material is impervious. This remains to be confirmed. For instance, one might use a similar approach as the one presented in the present paper but for a system where the pervaporation does not take place. In this respect, it can be noted that deformation of a porous material due to the drying of capillary bridges is reported in [28]. Although the situation in [28] is different from the one studied in the present paper, this is an indication that deformation at pore scale due to capillary effects are possible with an impervious solid matrix. In addition to shedding light on the stress generation problem, a major issue in relation with the degradation mechanism of porous materials due to salt, the present work has also confirmed several important points in the analysis of crystal growth in pores. Firstly, the classical expression of the crystallization pressure leads to estimate values consistent with our experiments. In other words, we have no particular reason to question the validity of Eq.(1) from our results. Secondly, our experiments can be also seen as a confirmation that a thin liquid does exist between the crystal and the wall. The existence of this film was essentially postulated from theoretical considerations in previous works, i.e. [5]. Our experimental observations and associated analyses are fully consistent with the existence of the film (see also the recent work [22] where the film thickness between a crystal in a solution and a glass plate could be measured using an interference method). Finally, our experiments have also led to identify and analyze a new phenomenon, the hyperslow drying process of PDMS channels.

REFERENCES

[1] I.N. Nassar, T. Horton, Salinity and compaction effects on soil water evaporation and water and solute distributions *Soil Sc. Soc. Am. J.* 63(3), 752 -758 . (1999)

[2] X.Y.Chen, Evaporation from a salt-encrusted sediment surface: Field and laboratory studies, Aust. J. Soil Res., 30(4), 429– 442 (1992)

[3] H. Eloukabi, N.Sghaier, S. Ben Nasrallah, M. Prat, Experimental study of the effect of sodium chloride on drying of porous media: the crusty-patchy efflorescence transition, Int. J. of Heat and Mass Tr., 56, 80–93, (2013)

[4] A.Goudie, H. Viles, Salt Weathering Hazards. Wiley, Chichester (1997).

[5] G.W. Scherer, Stress from crystallization of salt, Cem. Concr. Res. 34, 1613–1624 (2004).

[6] RJ Flatt, F Caruso, AMA Sanchez, GW Scherer, Chemo-mechanics of salt damage in stone, Nature communications 5, 4823 (2014)

[7] M.Schiro, E. Ruiz-Agudo, C. Rodriguez-Navarro, Damage Mechanisms of Porous Materials due to In-Pore Salt Crystallization, Phys. Rev. Lett. 109, 265503 (2012)

[8] C.Rodriguez-Navarro, E.C.Doehne, Salt weathering: influence of evaporation rate, supersaturation and crystallization pattern, Earth Surf. Processes Landforms, **24**(3) (1999) 191-209 (1999).

[9] M. Steiger, Crystal growth in porous materials—I: the crystallization pressure of large crystals, J.Cryst.Growth 282 (3–4), 455–469 (2005).

[10] J Desarnaud, D Bonn, N Shahidzadeh, The pressure induced by salt crystallization in confinement, Scientific Reports 6, 30856 (2016).

[11] A.Naillon, P.Joseph, M.Prat, Ion transport and precipitation kinetics as key aspects of the stress generation on pore walls induced by salt crystallization, Phys. Rev. Letters 120 (3), 034502 (2018)

[12] J. Desarnaud, H.Derluyn, J.Carmeliet, D.Bonn, N.Shahidzadeh, Metastability limit for the nucleation of NaCl crystals in confinement, J.Phys.Chem.Lett.5 (5) 890–895 (2014).

[13] A.Naillon, P.Duru, M.Marcoux, M.Prat, Evaporation with sodium chloride crystallization in a capillary tube, J. of Crystal Growth, 422, 52-61 (2015).

[14] A. Naillon, P. Joseph and M. Prat, Sodium chloride precipitation reaction coefficient from crystallization experiment in a microfluidic device, J. of Crystal Growth 463, 201-210 (2017)

[15] J. Stefan, Uber das gleichgewicht und die bewegung in besondere die diffusion von gasgemengen. Sitzungsber Math-Naturwiss. Akad. Wiss. Wien 63, 63–124 (1871).

[16] GC Randall, PS Doyle, Permeation-driven flow in poly (dimethylsiloxane) microfluidic devices, PNAS, 102 (31) 10813-10818 (2005).

[17] J.Leng, B.Lonetti, P.Tabeling, M.Joanicot, A.Ajdari, Microevaporators for Kinetic Exploration of Phase Diagrams, Phys. Rev. Letters 96, 084503 (2006)

[18] MP Milner, L Jin, SB Hutchens, Creasing in evaporation-driven cavity collapse, Soft Matter, 13, 6894-6904 (2017).

[19] A.Merlin, J.B. Salmon, J. Leng, Microfluidic-assisted growth of colloidal crystals, Soft Matter, 8, 3526-3537 (2012).

[20] S.J. Harley, E. A. Glascoe, R. S. Maxwell, Thermodynamic study on dynamic water vapor sorption in Sylgard-184, J. Phys. Chem. B, 116, 48, 14183-14190 (2012).

[21] G.W. Scherer, Stress from crystallization of salt in pores, in: V. Fassina (Ed) Proc. 9th Int. Cong. Deterioration and Conservation of Stones, Elsevier, Amsterdam, 2000, pp. 187–194.

[22] F. Kohler, L. Gagliardi, O. Pierre-Louis, D. K. Dysthe, Cavity formation in confined growing crystals, Phys. Rev. Letters 121, 096101 (2018).

[23] J.W. Mullin, Crystallization, 4th ed. (Butterworth Heinemann, Oxford, 2001).

[24] JN Israelachvili, Intermolecular and surface forces, 3th ed, Academic Press (2011).

[25] T.Corti, U.K. Krieger, Improved inverted bubble method for measuring small contact angles at crystal-solution-vapor interfaces. Applied Optics, 46(23):5835 (2007).
[26] L. Gagliardi, O. Pierre-Louis, Crystal growth in nano-confinement: subcritical cavity formation and viscosity effects, New J. Phys. 20, 073050 (2018).
[27] X.Noblin, NO Rojas, J Westbrook, C. Llorens, M Argentina, J Dumais, The fern sporangium: a unique catapult, Science 335 (6074), 1322-1322 (2012).
[28] M. Bouzid, L. Mercury, A. Lassin, J. M. Matray, M. Azaroual, In-pore tensile stress by drying-induced capillary bridges inside porous materials, J. Colloid Interface Sci., 355, 494–502 (2011).

# Supplementary Material

for the manuscript

# Direct observation of pore collapse and tensile stress generation on pore wall due to salt crystallization

A.Naillon[1,2], P.Joseph[2], M.Prat[1*]

[1]*Institut de Mécanique des Fluides de Toulouse (IMFT), Université de Toulouse, CNRS, Toulouse, France*
[2]*LAAS-CNRS, Université de Toulouse, CNRS, Toulouse, France*

## A. Experimental set-up

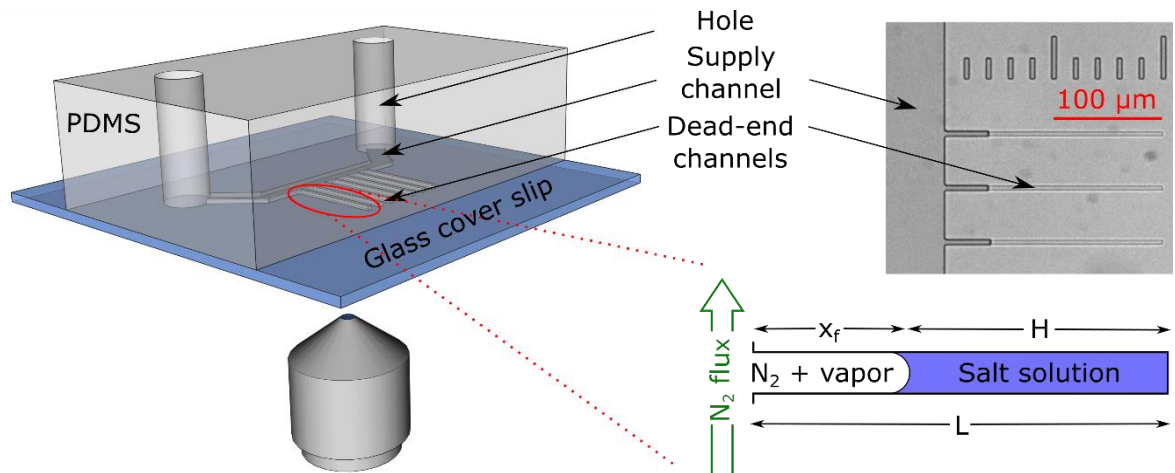

**FIG. A1.** (Color online) Schematic of the PDMS and glass microfluidic chip. Crystallization and wall deformations are observed in the dead-end channels.

The experimental set-up is composed of a large channel used for supplying the fluids: salt solution or gaseous nitrogen. 200 µm long smaller channels of 5×5 µm² square cross-section, referred to as pore channels, are positioned perpendicularly to the supply channel. Details on the microfluidic chip fabrication procedure are given in [11]. Note that estimates from the images suggest that the actual width is rather 4.5 µm. The latter value is adopted in what follows. The crystallization is triggered by evaporation of the sodium chloride solution confined in the pore channels. Salt solution is provided from the top hole through the supply channel and invades the pore channels. Once the device is filled, a dry $N_2$ flux is imposed from the bottom hole to empty the supply channel and isolates salt solution in the pore channels. This flux is maintained during all the experiment. As a result of evaporation and pervaporation through the PDMS, the meniscus recedes into the pore channel, the ion mass fraction increases until the ion mass fraction $C_{cr}$ marking the onset of crystallization is reached. This leads to the formation of a single crystal, most often within the liquid bulk away from the receding meniscus.

---

* Corresponding author : mprat@imft.fr, +33 (0)5 34 32 28 83

## B. Hyperslow drying in PDMS channel (Fig.3 in main text)

The situation under consideration is sketched in Fig.B1.

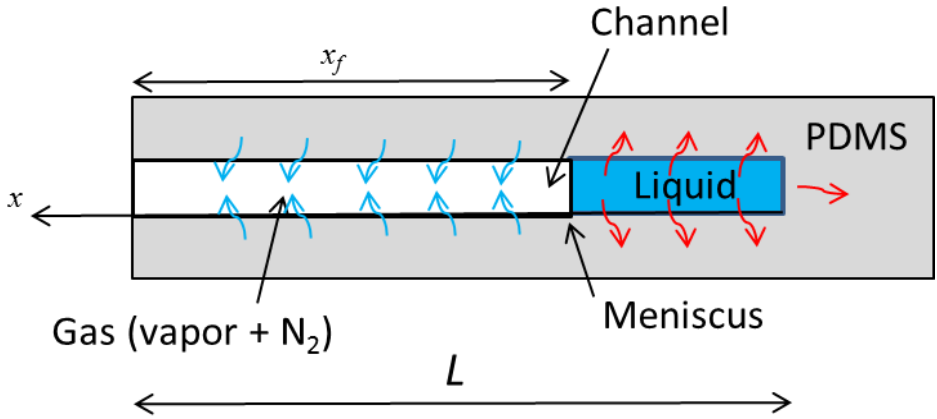

**Fig.B1.** Sketch of the considered situation. The blue arrows represent the water mass transfer from water saturated PDMS toward the section of the channel occupied by the gas phase. The red arrows represent the pervaporation process.

Let $v_{pe}$ be the pervaporation velocity at the channel PDMS wall in the liquid plug. For simplicity, $v_{pe}$ is assumed constant and uniform over the channel PDMS wall. Then a simple mass balance is expressed as

$$\rho W^2 \frac{dx_f}{dt} = \rho 3W(L - x_f)v_{pe} - J \qquad (B1)$$

where $W$ is the channel width and the factor 3 comes from the fact that the channel has three PDMS wall (there is no pervaporation from the fourth one in glass). $J$ is the "condensation" flux at the moving meniscus. The existence of $J$ is explained as follows. If one considers that the vapor concentration at channel wall in the gaseous part of the channel (at least in the vicinity of the receding meniscus) is the equilibrium vapor concentration for pure water [20] (since the ions do not penetrate PDMS) and that the vapor concentration at the receding meniscus is less (the equilibrium vapor concentration of a NaCl saturated solution is 25% less than the equilibrium vapor concentration for pure water. When the solution is supersaturated, the vapor concentration at the meniscus can be even lower), then a water transfer must occur between the PDMS wall and the receding meniscus by diffusion in the gas phase.

This "condensation" mechanism should contribute to slow down the meniscus. Considering *J* as a constant leads to a very good agreement with the experimental data.

Solving Eq.(B1) is straightforward. The solution reads

$$x_f = L - \left(\frac{J}{3\rho W v_{pe}} - \left(\frac{J}{3\rho W v_{pe}} - L\right)\exp\left(-\frac{3v_{pe}t}{W}\right)\right) \quad (B2)$$

The pervaporation velocity is estimated in Appendix E ($v_{pe}$ = 2.3 x $10^{-8}$ m/s). Using this value in Eq.(B2) with the dimension of the channel ($W \approx$ 4.5 µm, $L$ = 200 µm) and $J$ = 3.35 $10^{-14}$ kg/s leads to the very good agreement with the experimental data shown in Fig.3b (main text). However, it should be clear that the value of *J* has been adjusted to get this excellent agreement. A more comprehensive analysis would imply to predict *J* from the modelling of the coupled transport phenomena between the PDMS walls and the other regions (supply channel, pore channel, external air). This is left for a future study which will probably require some numerical simulations.

Also, as mentioned in the main text, the fact that the walls are humid (due to the presence of water in PDMS) in the gaseous part of the channel explains why the classical diffusion controlled evaporation in a tube (Stefan's tube situation) is negligible in the case of our experiment.

## C. Meniscus acceleration induced by the crystal growth (Fig.2b in main text)

The situation analyzed in this sub-section is sketched in Fig.C1.

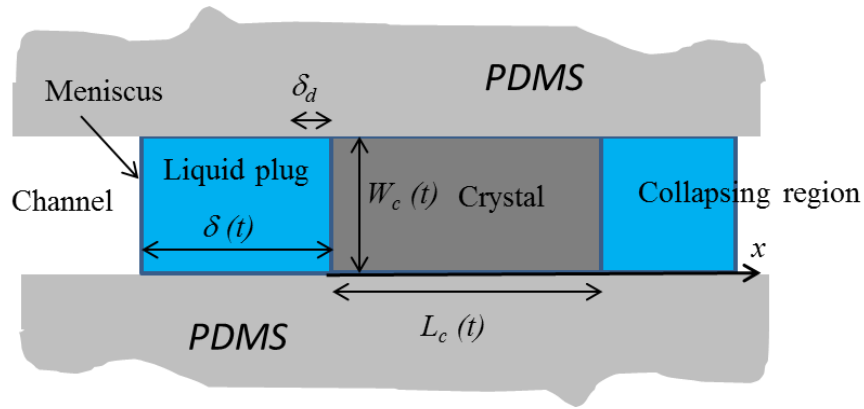

**Fig.C1**. *Schematic of considered situation.*

The objective is to explain the sudden acceleration of the meniscus during the crystal growth depicted in Fig.2b (strong increase in the slope of the curve showing the variation of the meniscus position as a function of time in Fig.2b). Referring to Fig.C1, The objective is thus to analyze the variation of the liquid plug length $\delta(t)$.

For convenience, the rapid variation of the crystal back face width $W_c(t)$ shown in Fig.2a is represented by a third degree polynomial,

$$W_c(t) = -8869.5 + 72.334\,t - 0.19648\,t^2 + 0.00017787\,t^3 \quad \text{for } 348\text{ s} \leq t \leq 364\text{ s} \tag{C1}$$

where $t$ is the time in seconds and $W_c(t)$ is in $\mu m$.

The meniscus sudden acceleration is analyzed from the consideration of two effects, the channel expansion effect and the crystallization induced flow. The two effects are both taken into account to obtain the result shown in Fig.2b (inset). For simplicity, we begin with the consideration of each effect separately.

*Channel expansion effect*
This effect refers to the fact that the conservation of the liquid plug mass implies that the liquid plug length $\delta(t)$ must decrease in the channel when the width increases due to the growth of the crystal. The volume of the liquid plug on the left of the crystal in Fig.C1 is expressed as

$$V = \delta W_c^2 \tag{C2}$$

Since the duration of the meniscus acceleration period (~10 s) is small compared to the collapse period (~ 100 s), it is assumed that the mass loss due to pervaporation can be neglected. In other words, it is assumed that the volume $V$ does not vary significantly. As a result,

$$\delta W_c^2 = \delta_0 W_{c0}^2 \tag{C3}$$

where the subscript « 0 » refers to values at the very beginning of the meniscus acceleration period. Thus,

$$\delta = \delta_0 \frac{W_{c0}^2}{W_c^2} \tag{C4}$$

*Liquid flow induced by the crystallization*
This effect refers to the fact that the crystal growth induced a flow directed on average toward the crystal in the adjacent liquid. As presented in [14], the kinematic condition at the crystal liquid interface reads

$$\boldsymbol{v}_l \cdot \boldsymbol{n}_{cr} = \left(1 - \frac{\rho_{cr}}{\rho_l}\right) \boldsymbol{w}_{cr} \cdot \boldsymbol{n}_{cr} \tag{C5}$$

where $\boldsymbol{w}_{cr}$ is the velocity of the crystal-solution interface, $\boldsymbol{v}_l$ is the liquid velocity at the crystal –liquid interface, $\boldsymbol{n}_{cr}$ is the unit normal vector at the interface, $\rho_{cr}$ is the crystal density (2160 kg/m$^3$), $\rho_l$ is the solution density (~1200 kg/m$^3$). Based on the results shown in Fig.2, it is assumed that the crystal growth essentially occurs over the four faces of the crystal parallel to the channel wall during the very short period when the meniscus acceleration occurs (on the ground that the growth of the crystal faces perpendicular to the channel wall is quite weak during the considered period). Accordingly, the total flow rate induced in the liquid is estimated as

$$Q(t) \approx \left(\frac{\rho_{cr}}{\rho_l} - 1\right) 4\, L_c W_c(t) \left(\frac{1}{2}\frac{dW_c(t)}{dt}\right) \tag{C6}$$

Then expressing that this flow rate should correspond to the meniscus displacement leads to

$$W_c^2 \frac{d\delta}{dt} = -Q(t) \approx -2\left(\frac{\rho_{cr}}{\rho_l} - 1\right) L_c W_c(t) \frac{dW_c(t)}{dt} \tag{C7}$$

*Combining both effects*
Both effects can be taken into account as follows. From Eq.(C2) and taking into account the flow rate induced by the salt precipitation yields,

$$\frac{dV}{dt} = \frac{d(\delta W_c^2)}{dt} = -Q(t) \tag{C8}$$

which can be expressed as

$$W_c^2 \frac{d\delta}{dt} + 2\delta W_c \frac{dW_c}{dt} = -Q(t) \tag{C9}$$

Then, taking into account Eq.(C6) leads to express Eq.(C9) as

$$W_c^2 \frac{d\delta}{dt} + 2\delta W_c \frac{dW_c}{dt} = -2\left(\frac{\rho_{cr}}{\rho_l} - 1\right) L_c W_c(t) \frac{dW_c(t)}{dt} \tag{C10}$$

or

$$W_c^2 \frac{d\delta}{dt} = -\left(2\delta W_c + 2\left(\frac{\rho_{cr}}{\rho_l} - 1\right) L_c W_c(t)\right) \frac{dW_c}{dt} \tag{C11}$$

Actually, the deformation of the channel occurs in the direction of the liquid plug over a distance which is smaller than the initial length $\delta_0$ of the liquid plug. In other words, it is assumed that the channel deformation in the liquid plug region occurs over a distance $\delta_d$ from the crystal (with $\delta_d < \delta_0$).
Under these circumstances, the volume of the liquid plug can be expressed as,

$$V = (\delta - \delta_d)W_0^2 + \delta_d W_c^2 \quad \text{when } \delta \geq \delta_d \tag{C12}$$

$$V = \delta W_c^2 \quad \text{when } \delta \leq \delta_d \tag{C13}$$

Then the liquid plug mass conservation equation can be expressed as follows when $\delta \geq \delta_d$,

$$\frac{dV}{dt} = \frac{d\left((\delta - \delta_d)W_0^2 + \delta_d W_c^2\right)}{dt} = -Q(t) \tag{C14}$$

leading to

$$W_0^2 \frac{d\delta}{dt} + 2\delta_d W_c \frac{dW_c}{dt} = -Q(t) \tag{C15}$$

Substituting Eq.(C6) into Eq.(C15) leads to

$$W_0^2 \frac{d\delta}{dt} + 2\delta_d W_c \frac{dW_c}{dt} = -2\left(\frac{\rho_{cr}}{\rho_l} - 1\right) L_c W_c(t) \frac{dW_c(t)}{dt} \tag{C16}$$

$$W_0^2 \frac{d\delta}{dt} = -\left(2\delta_d W_c + 2\left(\frac{\rho_{cr}}{\rho_l} - 1\right) L_c W_c(t)\right) \frac{dW_c(t)}{dt} \tag{C17}$$

Eq.(C17) is used as long as $\delta \geq \delta_d$. When $\delta \leq \delta_d$, then one can use Eq.(C11). With $\delta_d = 2$ μm, which seems to be a reasonable value, using Eq.(C17) and Eq.(C11) together with Eq.(C1) leads to the results show in the inset of Fig.2b (where "model" corresponds to the numerical values obtained from Eq.(C17) and Eq.(C11)).

## D. Growth of crystal front face (Fig.4 in main text)

We consider the situation sketched in Fig.D1

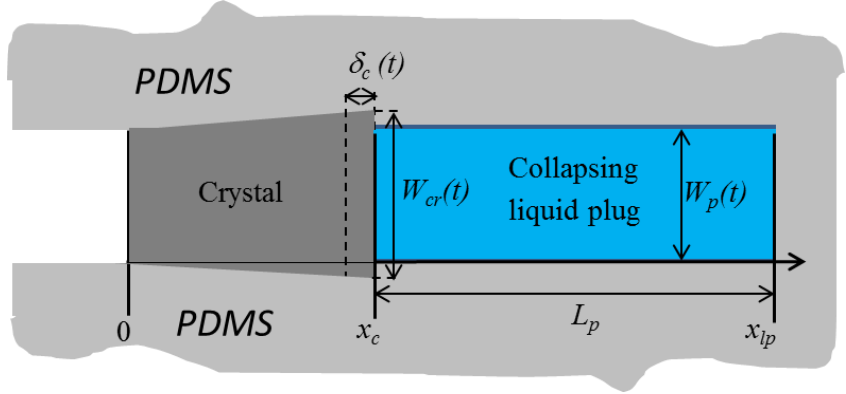

**Fig.D1.** *Schematic of considered situation.*

Let $V$ be the volume of the collapsing liquid plug. Assuming that the ion mass fraction in the plug is very close to the equilibrium mass fraction $C_{eq}$ on the ground that the NaCl precipitation reaction is quite fast, e.g. [14], the initial mass of salt in the plug is expressed as

$$m_0 = \rho_\ell C_{sat} V_0 \tag{D1}$$

where we have assumed that $C_{eq} \approx C_{sat}$ ($C_{sat}$ is the solubility in the reference state).
The mass of salt in solution at time $t$ is

$$m = \rho_\ell C_{sat} V(t) \tag{D2}$$

where $\rho_\ell$ is the solution density.
Then the mass flow rate of salt crystallizing is

$$\phi_s = \frac{dm}{dt} = \rho_\ell C_{sat} \frac{dV(t)}{dt} \tag{D3}$$

Denoting by $W_{cr0}$ the size of the crystal when the collapse begins, Eq.(D3) can be expressed as

$$W_{cr0}^2 \rho_{cr} \frac{d\delta_c(t)}{dt} = -\rho_\ell C_{sat} \frac{dV(t)}{dt} \tag{D4}$$

where $\delta_c$ is the increase in the length of the crystal on the right (see Fig.D1) and $\rho_{cr}$ is the crystal density. This yields

$$\frac{d\delta_c(t)}{dt} = -\frac{\rho_\ell C_{sat}}{W_{cr0}^2 \rho_{cr}} \frac{dV(t)}{dt} \tag{D5}$$

For simplicity we express $V(t)$ as $V(t) = W_p^2(t) L_p(t)$ where $L_p$ is the length of the plug ($L_p$ slightly varies owing to the crystal growth in the direction of channel dead end) and $W_p$ is the width of the plug. Then Eq.(D5) can be expressed as

$$\frac{d\delta_c(t)}{dt} = -\frac{\rho_\ell C_{sat}}{W_{cr0}^2 \rho_{cr}} \frac{dW_p^2 L_p}{dt} \tag{D6}$$

This leads to

$$\frac{d\delta(t)}{dt} = -2W_p(t) L_p(t) \frac{\rho_\ell C_{sat}}{W_{cr0}^2 \rho_{cr}} \frac{dW_p(t)}{dt} - W_p^2(t) \frac{\rho_\ell C_{sat}}{W_{cr0}^2 \rho_{cr}} \frac{dL_p(t)}{dt} \tag{D7}$$

A fit of the experimental results for $W_p$ shown in Fig. 2 (collapsing channel width) gives

$$W_p(t) = -15.46 + 0.14294t - 0.000224405t^2 \qquad \text{for } 360 \text{ s} \leq t \leq 440 \text{ s} \tag{D8}$$

The plug is initially about 22 μm long. Thus $L_{p0} = 22$ μm and

$$L_p(t) \approx L_{p0} - \delta_c(t) \tag{D9}$$

which leads to express Eq.(D7) as

$$\frac{d\delta_c(t)}{dt} = -2W_p(t)(L_{p0} - \delta_c(t)) \frac{\rho_\ell C_{sat}}{W_{cr0}^2 \rho_{cr}} \frac{dW_p(t)}{dt} + W_p^2(t) \frac{\rho_\ell C_{sat}}{W_{cr0}^2 \rho_{cr}} \frac{d\delta_c(t)}{dt} \tag{D10}$$

thus

$$\left(1 - W_p^2(t) \frac{\rho_\ell C_{sat}}{W_{cr0}^2 \rho_{cr}}\right) \frac{d\delta_c(t)}{dt} = -2W_p(t)(L_{p0} - \delta_c(t)) \frac{\rho_\ell C_{sat}}{W_{cr0}^2 \rho_{cr}} \frac{dW_p(t)}{dt} \tag{D11}$$

The collapsing channel cross section shape is expected to be somewhat different from a square shape since the glass cover plate does not deform and the deformation in the channel corner region should be less than in the middle of the channel walls. In other words, it can be argued that the cross section area of the collapsing channel is greater than $W_p^2$. We introduce a shape factor $F$ for taking into account this effect, $W_{peff} = F\,W_p$. This leads to express Eq.(D11) as,

$$\left(1 - 2\,F^2 W_p^2(t)\,\frac{\rho_\ell C_{sat}}{W_{cr0}^2 \rho_{cr}}\right)\frac{d\delta_c(t)}{dt} = -2F^2 W_p(t)(L_{p0} - \delta_c(t))\,\frac{\rho_\ell C_{sat}}{W_{cr0}^2 \rho_{cr}}\frac{dW_p(t)}{dt} \tag{D12}$$

With $L_{p0} = 22$ μm, $\rho_\ell \approx 1200$ kg/m³, $\rho_{cr} = 2160$ kg/m³, $C_{sat} = 0.264$, $W_{ce0} \approx 6.2$ μm, solving numerically Eq.(D12) gives the results shown in Fig.4. As can been, a quite reasonable agreement is obtained with the experiment with $F = 1.22$.

**E. Estimate of pervaporation velocity**

We consider the situation sketched in Fig.D1.

The pervaporation velocity $v_{pe}$ is defined as

$$v_{pe} = \frac{J_{pe}}{A_{pe}\rho_\ell} \tag{E1}$$

where $\rho_\ell$ is the density of the solution, $J_{pe}$ is the pervaporation rate and $A_{pe}$ is the surface area of the PDMS walls limiting the collapsing region. Thus, $v_{pe}$ is the velocity perpendicular to the wall induced in the solution by the pervaporation process.

The mass of solution in the collapsing region is expressed as

$$m_l = \rho_\ell W_p^2 L_p \tag{E2}$$

Where $W_p$ is the width of the collapsing region and $L_p$ is the length of the collapsing region (see Fig.D1).

The solution mass balance in the collapsing region is expressed as

$$\frac{dm_l}{dt} = J_{pe} + J_{cr} \tag{E3}$$

where $J_{cr}$ is the mass flow rate resulting from the longitudinal growth of the crystal inside the collapsing region. The mass balance at the moving crystal front face reads [14],

$$J_{cr} = -\rho_{cr} w_{cr}\,W_{cr}^2 \tag{E4}$$

where $w_{cr}$ is the velocity of the crystal front face, $W_{cr}$ is the width of the crystal front face (see Fig.D1) and $\rho_{cr}$ is the crystal density. Eq.(E4) can be expressed as

$$J_{cr} = \rho_{cr}\,W_{cr}^2\,\frac{dL_p}{dt} \tag{E5}$$

Combining the above equations and noting that $A_{pe} = 3 L_p W_p + W_p^2$, where the factor 3 comes from the fact that the pervaporation takes place only through three walls of the channel (no pervaporation through the glass plate, see Fig.E1) and the factor $W_p^2$ corresponds to the surface of the channel tip, leads to the following expression of the pervaporation velocity:

$$v_{pe} = -\frac{\left[2L_p W_p \frac{dW_p}{dt} + \left(W_p^2 - \frac{\rho_{cr}}{\rho_\ell} W_{cr}^2\right)\frac{dL_p}{dt}\right]}{3L_p W_p + W_p^2} \tag{E6}$$

$\frac{dW_p}{dt}$ and $\frac{dL_p}{dt}$ are estimated from linear fits of the experimental data over the time period $362s \leq t \leq 392s$ corresponding to the initial period of collapse when the channel is not yet too deformed. This gives $\frac{dW_p}{dt} \approx -0.042$ μm/s and $\frac{dL_p}{dt} \approx -0.037$ μm/s. With $L_p \approx 22$ μm, $W_p \approx 4.5$ μm, $W_{cr} = 6.5$ μm (Fig.2a), $\rho_{cr} = 2160$ kg/m³, $\rho_\ell = 1200$ kg/m³, one obtains from Eq.(E6), $v_{pe} = 2.3 \times 10^{-8}$ m/s. This corresponds to a pervaporation flux $j_{pe} = \rho_\ell v_{pe} \approx 2.8 \times 10^{-5}$ kg/m²/s.

Interestingly, this estimate is consistent with the estimate that can be obtained from the formula used in [16]. This formula reads

$$j_{pe} = -\frac{\pi D_p \rho_{sat}}{W_p \ln\left(\frac{W_p}{4R}\right)} \tag{E7}$$

where $D_p$ ($D_p \approx 8 \times 10^{-10}$ m²/s) is the diffusion coefficient for water in PDMS, $\rho_{sat}$ is the saturation water concentration in PDMS ($\rho_{sat} = 0.72$ kg/m³). The geometry considered in [16] is the one of a very small channel in the middle of a hemi-cylindrical PDMS domain of radius $R$. With the approximation that $R$ is about equal to the thickness of the PDMS layer (5 mm), using Eq.(E7) yields $j_{pe} = 4.4 \times 10^{-5}$ kg/m²/s. This value is quite close to the one estimate above and thus is considered as a confirmation that the pervaporation process controls the collapse kinetics. The slightly lower value can be due to the activity of the solution which is less than pure water as well as the humidity in the external air which is not zero in our experiments.

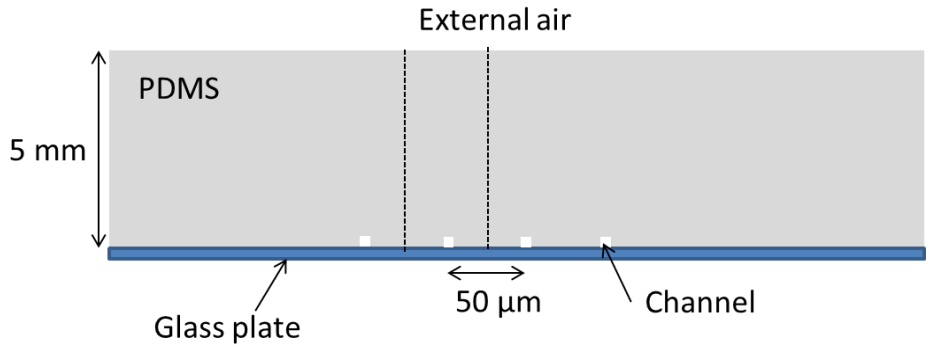

**Fig. E1**. Schematic of the experimental device cross-section.

## F. Ion transport distribution in the thin film (Fig.6 in main text)

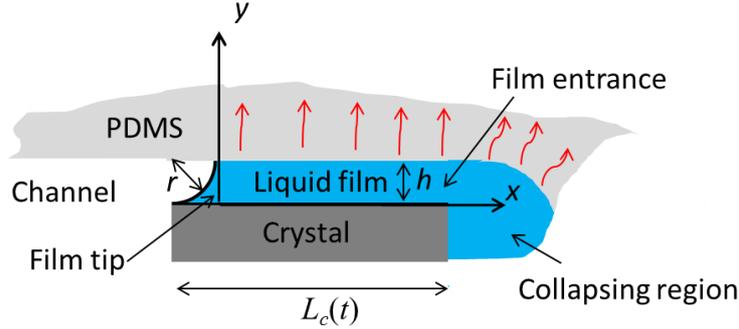

**Fig.F1.** *Schematic of thin film confined between crystal and PDMS channel wall. The red arrows represent the pervaporation process. The figure is not at scale. The film thickness h is expected to be on the order of* 10-100 nm *whereas the crystal length $L_c \approx$ 10 μm.*

The 1d version of the ion transport equation in the film reads,

$$\frac{\partial C}{\partial t} + \frac{\partial}{\partial x}(vC) = D\frac{\partial^2 C}{\partial x^2} - a_v k_r (C - C_{eq}) \tag{F1}$$

where $C$ is the height–averaged ion mass fraction, $v$ is the height averaged velocity, $D$ is the diffusion coefficient of the ions in the solution, $k_r$ is the precipitation reaction coefficient, $a_v$ is the specific surface area ($a_v = \frac{W_c}{W_c h} = h^{-1}$, where $W_c$ is the width of the crystal), $C_{eq}$ is the equilibrium ion mass fraction in the solution.

From mass conservation the height-averaged velocity in the solution is given by

$$v = -x\frac{v_{pe}}{h} \tag{F2}$$

where $v_{pe}$ is the pervaporation velocity ($v_{pe} = j/\rho_\ell$, where $j$ is the pervaporation flux through the PDMS and $\rho_\ell$ is the density of the solution). The maximum velocity (in absolute value) is at the entrance of the film (at $x = L_c$ in Fig.F1).

$$v_{max} = -v_{pe}\frac{L}{h} \tag{F3}$$

Then, the Peclet number characterizing the competition between advective and diffusive transports along the film can be expressed as,

$$Pe = \frac{\|v_{max}\|L_c}{D} = \frac{v_{pe}L_c^2}{hD} \tag{F4}$$

From the computation of pervaporation velocity ($v_{pe} = 2.3 \times 10^{-8}$ m/s, see SI Appendix E), one gets with $L_c \approx 10$ μm, $h \approx 10\text{-}100$ nm, $D = 1.3 \times 10^{-9}$ m$^2$/s, $Pe \approx 0.03 - 0.3$. Based on the low value of the Peclet number, a reasonable simplification is to neglect the convective term in Eq.(F1),

$$\frac{\partial C}{\partial t} = D\frac{\partial^2 C}{\partial x^2} - a_v k_r (C - C_{eq}) \tag{F5}$$

A characteristic time for diffusion is $\frac{L_c^2}{D} \approx 0.1$ s. This time is short compared to the crystal growth time (O(10-100s) as shown in Fig.2 in the main text). Thus, the evolution of the ion mass fraction in the film can be considered as quasi-steady.

$$D\frac{\partial^2 C}{\partial x^2} - a_v k_r (C - C_{eq}) = 0 \tag{F6}$$

Eq. (F6) is associated with the following boundary conditions,

$$D\frac{\partial C}{\partial x} = 0 \quad \text{at } x = 0 \tag{F7}$$

$$C = C_{L_c} \quad \text{at } x = L_c(t) \tag{F8}$$

Eq.(F7) expresses that the ions cannot leave the film through the meniscus on the left whereas $C_{L_c}$ (Eq.(F8)) is the ion mass fraction at the entrance of the film. The latter is estimated from Eq.(2) (main text) and the longitudinal growth rate of the front face. From Eq.(2) and Fig.2b,
$w_{cr} = \frac{k_r \rho_\ell}{\rho_{cr}}(C_{L_c} - C_{eq}) = \frac{\Delta W}{\Delta t} \approx \frac{180.04 - 177.69}{402 - 348} = 4.35 \times 10^{-2}$ μm/s. With $\rho_{cr} = 2160$ kg/m$^3$, $\rho_\ell = 1200$ kg/m$^3$, and $k_r \sim 2.3 \cdot 10^{-2}$ m/s [14], one obtains $C_{L_c} - C_{eq} = 0.000034$.
$C^* = C - C_{eq}$. Eqs. (F6-F7) are expressed as

$$D\frac{\partial^2 C^*}{\partial x^2} - a_v k_r C^* = 0 \tag{F9}$$

$$D\frac{\partial C^*}{\partial x} = 0 \quad \text{at } x = 0 \tag{F10}$$

$$C^* = C_{L_c} - C_{eq} \quad \text{at } x = L_c(t) \tag{F11}$$

The solution of Eq.(F9) reads

$$C^* = C_1 \exp(\lambda x) + C_2 \exp(-\lambda x) \tag{F12}$$

where $\lambda = \sqrt{\frac{a_v k_r}{D}}$. After substitution in Eqs.(F10) and (F11), constants $C_1$ and $C_2$ are determined. the solution reads,

$$C^* = (C_{L_c} - C_{eq})\frac{(\exp(\lambda x) + \exp(-\lambda x))}{(\exp(\lambda L_c) - \exp(-\lambda L_c))} \tag{F13}$$

or

$$C = C_{eq} + (C_{L_c} - C_{eq}) \frac{(\exp(\lambda x) + \exp(-\lambda x))}{(\exp(\lambda L_c) - \exp(-\lambda L_c))} \tag{F14}$$

The length scale $1/\lambda$ is on the order of 100 nm, thus much smaller than the length of the crystal (~10 µm). As a result, the ion mass fraction is greater than $C_{eq}$ only over a small region at the entrance of the film (on the right). This is illustrated in Fig.6 in the main text where ion mass fraction profiles along the film given by Eq.(F14) are plotted.

It can be argued that we have considered that the value of the diffusion coefficient in the thin film was the same as in a non-confined liquid. Actually, discussions with experts and a short look at literature indicate that the confinement must be much more severe ($h$ ~1 nm) for expecting a noticeable impact of confinement of ion diffusion properties in the film.

**G. Estimate of collapsing pressure**

Collapsing pressure is estimated from simulations performed with Comsol Multiphysics 5.2©, a commercial software based on the finite element method. Because of the elongated geometry of the channel, we assume that the problem can be simplified as a plane stress problem. Thus, the simulations are performed in 2 dimensions. Moreover, the deformation of PDMS and glass are assumed to be linearly elastic. The mechanical model is based on Hooke's law applied to both materials, i.e. glass and PDMS:

$$\boldsymbol{\sigma} = \frac{E}{1+\upsilon}\left(\boldsymbol{\varepsilon} + \frac{\upsilon}{1-2\upsilon}\text{Tr}(\boldsymbol{\varepsilon})\mathbf{I}\right), \tag{G1}$$

where $\boldsymbol{\sigma}$, $\boldsymbol{\varepsilon}$ and $\mathbf{I}$, are the stress tensor, the strain tensor and the identity tensor, respectively. $E$ (MPa) and $\upsilon$ are the young modulus of the considered material and its Poisson coefficient with $E_{PDMS}$=1.2 MPa, $E_{glass}$=64 Gpa, $\upsilon_{PDMS}$=0.45 and $\upsilon_{glass}$=0.45.

On channel walls, a pressure load is imposed as boundary condition:

$$\sigma.\boldsymbol{n} = -P_s.\boldsymbol{n}, \tag{G2}$$

where $\boldsymbol{n}$ is the wall normal unit vector.

Fig.G1 shows results for a homogeneous normal stress of 0.3 MPa. Computation is not performed for higher (negative) pressure because the mesh distorted too much to be stable. From the results obtained in the range of normal stress [0, 0.3 MPa], it is inferred that a value of 0.5-0.6 MPa is a good order of magnitude of the pressure needed to collapse the channel in the middle of the collapsing region.

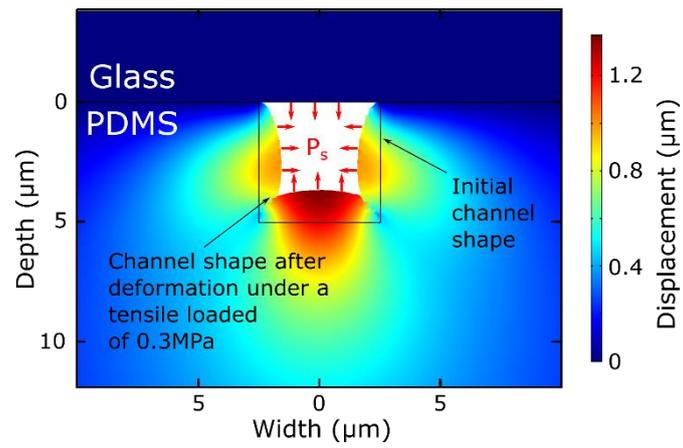

**Fig. G1**. Computation of the pore channel deformation under a homogeneous mechanical tensile load of 0.3MPa.